\documentclass{amsart}

\usepackage{graphicx}

\usepackage{xcolor}

\theoremstyle{definition}

\theoremstyle{remark}

\numberwithin{equation}{section}

\newcommand{\ue}{\mathrm{e}}
\newcommand{\ui}{\mathrm{i}}
\newcommand{\X}{\mathcal{X}}

\newcommand{\pd}[2]{\frac{\partial #1}{\partial\/ #2}}

\begin{document}

\title[On some model equations for pulsatile flow in viscoelastic vessels]{On some model equations for pulsatile flow in viscoelastic vessels}


\author{Dimitrios Mitsotakis}
\address{\textbf{D.~Mitsotakis:} Victoria University of Wellington, School of Mathematics and Statistics, PO Box 600, Wellington 6140, New Zealand}
\email{dimitrios.mitsotakis@vuw.ac.nz}

\author{Denys Dutykh}
\address{\textbf{D.~Dutykh:} Univ. Grenoble Alpes, Univ. Savoie Mont Blanc, CNRS, LAMA, 73000 Chamb\'ery, France}
\email{Denys.Dutykh@univ-smb.fr}

\author{Qian Li}
\address{\textbf{Q.~Li:} Victoria University of Wellington, School of Mathematics and Statistics, PO Box 600, Wellington 6140, New Zealand}
\email{ liqian5@myvuw.ac.nz}

\author{Elijah Peach}
\address{\textbf{E.~Peach:} Victoria University of Wellington, School of Mathematics and Statistics, PO Box 600, Wellington 6140, New Zealand}
\email{ elijah.peach@gmail.com}




\date{\today}



\begin{abstract}
Considered here is the derivation of partial differential equations arising in pulsatile flow in pipes with viscoelastic walls. The equations are asymptotic models describing the propagation of long-crested pulses in pipes with cylindrical symmetry. Additional effects due to viscous stresses in bio-fluids are also taken into account. The effects of viscoelasticity of the vessels on the propagation of solitary and periodic waves in a vessel of constant radius are being explored numerically.
\end{abstract}

\maketitle

\section{Introduction}

The description of fluid flows in pipes with viscoelastic wall material is motivated mainly by the studies of hemodynamics \cite{Fung2013}. A cardiac cycle consists of the systolic phase where the heart ventricles contract and pump blood to the arteries and the diastolic phase where the heart ventricles are relaxed and the heart fills with blood again. During the systolic phase the large arteries are deformed and store elastic energy that is released during the diastolic phase. This property of the vessels is usually referred to as the compliance of the vessels. Modelling the viscoelastic properties of the vessels appears to have significant difficulties of mathematical and numerical nature \cite{Fung2013, CYR2012, QF2004, VS2011,NOV2011,SPH2002}. The mathematical modelling of such flows suggests the use of the equations of continuum mechanics for incompressible fluid flow known as the Navier-Stokes equations.  

The Navier-Stokes equations in three dimensions are too complicated to be used in practical situations and for this reason several simplified mathematical models have been derived  \cite{Xiao2014,PR1995,THZ1998,Figueroa2006,BKKN2016,EED1992}. The main simplifications that have been made are based on the assumption of the axial (or cylindrical) symmetry of the vessels. This assumption and using approximations of the averaged velocity of the fluid led to the derivation of simple one-dimensional (1+1D) models \cite{HL1973,ZMV1986,TBMH1981,SYR1992,OPLPNL2000,Alastruey2011,RBPLS2011,WFL2014}. The models include unidirectional \cite{EED1992,Yomosa1987,MT2013,Demiray2007,Cascaval2003}, and bidirectional models \cite{BRV2007, MN2008,C2012}. Bidirectional models can approximate accurately reflections of pulses occurred in the presence of non-uniformities in the vessels wall as opposed to unidirectional models such as the KdV equation (\cite{Yomosa1987,Dem1996}) where it is assumed that the pulses propagate mainly in one direction. In this paper we focus on the derivation of bidirectional equations. Further simplifications were made by assuming that the velocity of the fluid is very large. This assumption led to lumped-parameter (or zero-dimensional) models \cite{APPS2008,MQ2004} and the references therein. The later models are used  in practice to predict the flow and the pressure of the blood in operational situations \cite{VS2011}. 

For practical reasons, the inclusion of the dissipative effects in the flow can be done by assuming a laminar flow and small viscosity. For example assuming a parabolic profile for the horizontal velocity of the fluid it has been shown that the Navier-Stokes equations can be reduced to a modified system which is very similar to the Euler equations \cite{SPH2002}, (we also refer to the Poiseuille solution for the justification of this parabolic profile of the horizontal velocity). Specifically, denoting $u=u(x,r,t)$, $v=v(x,r,t)$ the horizontal and radial velocity respectively, and $u^w(x,t)=u(x,r^w,t)$ the horizontal velocity of the fluid on the vessel wall (at radius $r=r^w(x,t)$), then assuming that  $u(x,r,t)$ is proportional to $(({r^w})^2-r^2)u^w(x,t)$ these equations can be written in cylindrical coordinates in the form:
\begin{align}
& u_t+uu_x+vu_r+\frac{1}{\rho}\;p_x+\kappa u^w=0\ , \label{eq:euler1}\\
& v_t+uv_x+vv_r+\frac{1}{\rho}\;p_r=0\ , \label{eq:euler2}\\
&u_x+v_r+\frac{1}{r}\;v=0 \ , \label{eq:euler3}
\end{align}
where  $p=p(x,r,t)$ is the pressure of the fluid, $\rho$ is the constant density of the fluid and $\kappa$ is the viscous frequency parameter (also known as the Rayleigh damping coefficient) with dimensions $[s^{-1}]$.

\begin{figure}
  \centering
  \includegraphics[width=0.8\columnwidth]{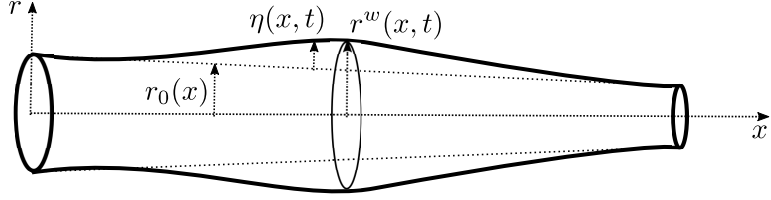}
  \caption{\small\em Sketch of the physical domain for a single vessel segment with elastic and impenetrable wall.}
  \label{fig:vessel}
\end{figure}

A sketch of the physical domain for this problem is presented in Fig. \ref{fig:vessel}, where the distance of vessel's wall from the centre of the vessel in a cross section is denoted by $r^w(x,t)$ and depends on $x$ and $t$ while the radius of the vessel at rest is given by the function $r_0(x)$. In general the deformation of the wall will be a function of $x$ and $t$.  If we denote the radial displacement of the wall by $\eta(x,t)$ then the vessel wall radius can be written as $r^w(x,t)=r_0(x)+\eta(x,t)$. 

The governing equations (\ref{eq:euler1})--(\ref{eq:euler3}) combined with initial and boundary conditions form a closed system. A compatibility condition is also applied at the centre of the vessel (due to cylindrical symmetry). Specifically, we assume that 
\begin{equation}\label{eq:bc1}
v(x,r,t)=0, \quad \mbox{ for }\quad r=0\ .
\end{equation}
In general we assume for consistency purposes that $v(x,r,t)=O(r^{1+\tau})$, with $\tau\geq 0$ as $r\rightarrow 0$.
The impermeability of the vessel wall can be described by the equation:
 \begin{equation}\label{eq:bc2}
 v(x,r,t)=\eta_t(x,t)+(r_0(x)+\eta(x,t))_x u(x,r,t), \mbox{ for } r = r^w(x,t)\ ,
 \end{equation}
and expresses that the fluid velocity equals the wall speed $v=r^w_t$.
The system is accompanied also by a second boundary condition, which is the Newton second law applied on the vessel wall:
\begin{equation}\label{eq:bc3}
\rho^w h \eta_{tt}(x,t)=p^w(x,t)-\frac{E_\sigma h}{r_0^2(x)}\;(\eta(x,t)+\gamma\eta_t(x,t))\ ,
\end{equation}
where $\rho^w$ is the wall density, $p^w$ is the transmural pressure, $h$ is the thickness of the vessel wall, $E_\sigma = E/(1-\sigma^2)$ where $E$ is the Young modulus of elasticity with $\sigma$ denoting the Poisson ratio of the viscoelastic wall. In this study we assume that $E$ is a constant and in general we simplify the notation by denoting $E_\sigma$ with $E$. The last term in (\ref{eq:bc3}) models the viscous nature of the vessel wall and can be derived by using a simple Kelvin-Voigt model (spring-dashpot model). In this setting $\gamma=n_w/E_v$ where $n_w$ is the dashpot coefficient of viscosity and $E_v$ is the Young modulus of the viscous part of Kelvin body. In practical situations the parameter $\gamma$ is very small and usually can be taken to be of order $O(10^{-4})$. It is noted that because the flow is pressure-driven the effect of gravity is neglected. For more information about the derivation of the Euler equations and the boundary conditions we refer to \cite{Zamir2000,CYR2012}. 

Although the dispersion of the flow can be ignored from the majority of mathematical models derived from the Navier-Stokes equations or from the Euler equations resulting into very simple systems of conservation laws, the need for more accurate description of the waves and their reflections suggests the inclusion of this fundamental property. A first attempt towards the derivation of bi-directional weakly-nonlinear and weakly-dispersive system of equations was presented in \cite{C2012}. Using asymptotic techniques more general asymptotic models were derived in \cite{MDL2017}. The systems derived in \cite{MDL2017} appeared to justify the non-dispersive models of \cite{FLQ2003,SFPP2003} with asymptotic reasoning. It was also shown that the inclusion of dispersive terms can describe more accurately the effects of the vessel wall variations within the flow. One basic ingredient that was ignored in both works \cite{C2012,MDL2017} is the viscosity effects of the vessels by assuming simple elastic vessels. 

In this paper we extend the work \cite{MDL2017} and derive some new asymptotic one-dimensional equations of Boussinesq type (weakly non-linear and weakly dispersive) that approximate the system (\ref{eq:euler1})--(\ref{eq:euler3}) with boundary conditions (\ref{eq:bc1})--(\ref{eq:bc3}). The derivation is based on formal expansions of the velocity potential as in \cite{DD2007}. The new systems generalise the previously derived Boussinesq systems of \cite{MDL2017} as they coincide with them when the viscoelastic property of the vessel wall is ignored. The new mathematical models are of significant importance since they include all the necessary ingredients for the accurate description of regular fluid flows in pipes with viscoelastic walls. 

Since the dissipation caused by the fluid viscosity and the dissipation caused by the viscoelasticity of the vessel wall are different in their nature there is a question on whether the different dissipative terms have also different effects on the propagation of pulses. The answer to this question is explored computationally by studying the effects caused by the two dissipative terms on the propagation of a solitary wave.

The paper is organised as follows: In Section \ref{sec:derivation} we present the derivation of a new system of Boussinesq type for the description of the velocity and the deviation of the vessel wall for fluid flow in a viscoelastic vessel. This system is further improved in Section \ref{sec:cbsys} by computing the fluid velocity at different levels of radius. In Section \ref{sec:unidir} further simplifications lead to unidirectional equations that depend only on the deviation of the vessel wall, while the velocity of the fluid can be computed explicitly using a simple asymptotic formula. Section 5 demonstrates the dissipation effects on the propagation of solitary and periodic waves. We close this paper with some conclusions and perspectives.

\section{Derivation of the new mathematical models}\label{sec:derivation}

Here we proceed with the derivation of the new equations. The derivation is based on the assumption that the flow is irrotational and therefore we assume the existence of a smooth velocity potential $\phi(x,r,t)$ such that $(u,v)^T=\nabla \phi$, i.e. we assume that $u=\phi_x$ and $v=\phi_r$. Then,  as in \cite{Whitham} the velocity potential can be chosen appropriately such that the equations (\ref{eq:euler1})--(\ref{eq:euler2}) can be integrated into the generalised Cauchy-Lagrange integral:
\begin{equation}\label{eq:bernouli}
\phi_t+\frac{1}{2}\;\phi^2_x+\frac{1}{2}\;\phi_r^2+\frac{1}{\rho}\;p +\kappa \phi=0, \quad \mbox{ for } r=r^w\ .
\end{equation}
The mass conservation (continuity) equation is then reduced to the elliptic equation
\begin{equation}\label{eq:massp}
r\phi_{xx}+(r\phi_r)_r=0, \quad \mbox 0<r<r^w\ ,
\end{equation}
and boundary conditions for the velocity are written as
\begin{equation}\label{eq:bc1p}
\phi_r=0, \mbox{ for } r=0\ ,
\end{equation}
and
 \begin{equation}\label{eq:bc2p}
 \phi_r=\eta_t+(r_0(x)+\eta)_x \phi_x, \mbox{ for } r = r^w(x,t)\ .
 \end{equation}

In order to make simplifications to the previous equations we consider the following non-dimensional (scaled) variables:
\begin{equation}\label{eq:ndv}
\eta^\ast=\frac{\eta}{a},~ x^{\ast}=\frac{x}{\lambda},~ r^{\ast}=\frac{r}{R},~  t^{\ast}=\frac{t}{T} \ ,\phi^\ast=\frac{1}{\lambda\varepsilon \tilde{c}}\phi,~ p^{\ast}=\frac{1}{\varepsilon \rho \tilde{c}^2}p\ , 
\end{equation}
where $a$ is a typical deviation of the vessel wall from its rest position, $\lambda$ a typical wavelength of a pulse, $R$ is a vessel's typical radius, $T=\lambda/\tilde{c}$ the characteristic time scale, while $\tilde{c}=\sqrt{Eh/2 \rho R}$ is the Moens-Korteweg characteristic speed, \cite{Fung2013}. It is noted that the external pressure is considered zero and is neglected. The parameters $\varepsilon$ and $\delta$ characterise the nonlinearity and the dispersion of the system:
\begin{equation}
\varepsilon = \frac{a}{R},\quad \delta = \frac{R}{\lambda}\ .
\end{equation}
Usually, $\varepsilon$ and $\delta$ are very small. Specifically, we assume that $\varepsilon\ll 1$, $\delta^2\ll 1$, while the Stokes-Ursell number is of order 1: $\varepsilon/\delta^2=O(1)$.  
The system of equations (\ref{eq:bernouli})--(\ref{eq:ndv}) along with the boundary condition (\ref{eq:bc3}) is then written in dimensionless variables in the form:
\begin{align}
& \phi^\ast_{t^\ast}+\frac{\varepsilon}{2}\;{\phi^\ast}^2_{x^\ast}+\frac{\varepsilon}{\delta^2}\;\frac{1}{2}\;{\phi^\ast}_{r^\ast}^2+p^\ast +\varepsilon\kappa^\ast \phi^\ast=0, \quad \mbox{ for } r^\ast={r^\ast}^w\ , \label{eq1}\\ 
& \delta^2 r^\ast\phi^\ast_{x^\ast x^\ast}+(r^\ast\phi^\ast_{r^\ast})_{r^\ast}=0, \quad \mbox 0<r^\ast<{r^\ast}^w\ , \label{eq2}\\
& \phi^\ast_{r^\ast}=0, \mbox{ for } r^\ast=0\ , \label{eq3}\\
&  \phi^\ast_{r^\ast}=\delta^2 \eta^\ast_{t^\ast}+\delta^2(r^\ast_0(x^\ast)+\varepsilon\eta^\ast)_{x^\ast} \phi^\ast_{x^\ast}, \mbox{ for } r^\ast = {r^\ast}^w\ , \label{eq4}\\
& p^\ast= \alpha^\ast \delta^2 \eta^\ast_{t^\ast t^\ast}+\beta^\ast(\eta^\ast+\delta^2\gamma^\ast \eta^\ast_{t^\ast}), \mbox{ for } r^\ast = {r^\ast}^w\ , \label{eq5}
\end{align}
where $\kappa^\ast=\kappa\lambda/\tilde{c}\varepsilon$, $\alpha^\ast=\rho^wh/\rho R$, $\beta^\ast=\beta^\ast(x)=2R^2/r_0^2(x)$ and $\gamma^\ast=\gamma/\delta^2T$.  For the sake of simplicity in the notation, we drop the asterisk from the new variables in the following derivations for the non-dimensional variables except if it is stated otherwise.

Following standard asymptotic techniques, cf. \cite{BCS2002}, we consider a formal expansion of the velocity potential \cite{Lagrange}:
\begin{equation}\label{eq:potential}
\phi(x,r,t)=\sum_{m=0}^\infty r^m \phi_m(x,t)\ .
\end{equation}
Demanding $\phi$ to satisfy the equation (\ref{eq2}) leads to the following recurrence relation
\begin{equation}\label{eq:recurr}
\delta^2\partial_x^2{\phi_{2m}}+(2m+2)^2 \phi_{2m+2}=0, \quad \phi_{2m+1}=0\ ,
\end{equation}
for $m=0,1,2,\cdots$, where $\partial_x^j$ denotes the $j$-th order derivative with respect to $x$. A direct application of the last relation is 
\begin{equation}\label{eq:phi2}
\phi_2=-\frac{\delta^2}{4}\;\partial_x^2{\phi_0}\ ,
\end{equation}
and
\begin{equation}\label{eq:phim}
\phi_{2m+2}=\frac{\delta^4}{(2m+2)^2(2m)^2}\;\partial_x^4\phi_{2m-2}=O(\delta^4)\ ,
\end{equation}
for $m=1,2,\cdots$.  The last relation ensures that the terms $\phi_m$ of the velocity potential expansion for $m\geq 4$ are negligible.
More general, we observe that
\begin{equation}\label{eq:phim2}
\phi_{2m}=(-1)^m\frac{\delta^{2m}}{2^{2m}}\;\partial^{2m}_x{\phi_0}\ ,
\end{equation}
for $m=1,2,\cdots$, and therefore
\begin{equation}\label{eq:pot1}
\phi(x,r,t)=\sum_{m=0}^\infty r^{2m}(-1)^m\frac{\delta^{2m}}{2^{2m}}\;\partial_x^{2m}\phi_0\ .
\end{equation}
A 2nd order asymptotic approximation of the velocity potential is
\begin{equation}\label{eq:potapr}
\phi(x,r,t)=\phi_0(x,t)-\delta^2\frac{r^2}{4}\;\partial_{xx}\phi_0(x,t)+O(\delta^4)\ .
\end{equation}

Using the previous observations on the expansion of the velocity potential we observe that (\ref{eq4}) can be approximated by the relation
\begin{equation}\label{eq:massp1}
\eta_t+r^w_x{\phi_0}_x+\frac{r^w}{2}\;{\phi_0}_{xx}-\delta^2 \frac{r_0^2{r_0}_x}{4}\;{\phi_0}_{xxx}-\delta^2\frac{r_0^3}{16}\;{\phi_0}_{xxxx}=O(\delta^4,\varepsilon\delta^2)\ .
\end{equation}

Since the momentum balance laws were reduced to the Cauchy-Lagrange integral equation (\ref{eq1}) for $r=r^w$ we can eliminate the pressure using (\ref{eq5}) and obtain
\begin{equation}\label{eq:bernoullinew}
\phi_t+\frac{\varepsilon}{2}\;\phi_x^2+\frac{\varepsilon}{\delta^2}\;\frac{1}{2}\;\phi_r^2+\varepsilon \kappa\phi+\alpha\delta^2\eta_{tt}+\beta(\eta+\gamma\eta_t)=0\ .
\end{equation}
Substituting (\ref{eq:potapr}) into (\ref{eq:bernoullinew}) we obtain the approximate momentum equation
\begin{equation}\label{eq:momentump1}
{\phi_0}_t-\delta^2\frac{r_0^2}{4}\;{\phi_0}_{xxt}+\frac{\varepsilon}{2}\;{\phi_0}_x^2+\varepsilon \kappa\phi_0+\alpha\delta^2\eta_{tt}+\beta(\eta+\delta^2\gamma\eta_t)=O(\delta^4,\varepsilon\delta^2)\ .
\end{equation}

Denoting the horizontal velocity at the centre of the vessel $u(x,0,t)={\phi_0}_x(x,t)$ by $w(x,t)$ we rewrite the equations (\ref{eq:massp1})--(\ref{eq:momentump1}) in the following form:
\begin{align}
& \eta_t+r^w_xw+\frac{r^w}{2}\;w_x-\delta^2 \frac{r_0^2{r_0}_x}{4}\;w_{xx}-\delta^2\frac{r_0^3}{16}\;w_{xxx}=O(\delta^4,\varepsilon\delta^2)\ , \label{eq:massp2}\\
& w_t+(\beta\eta)_x+\varepsilon w w_x+\varepsilon \kappa w-\delta^2\frac{r_0{r_0}_x}{2}\;w_{xt}-\delta^2\frac{r_0^2}{4}\;w_{xxt}+\alpha\delta^2\eta_{xtt}+\delta^2\gamma(\beta\eta_{t})_x=O(\delta^4,\varepsilon\delta^2)\ .\label{eq:momentump2}
\end{align}
Although the system (\ref{eq:massp2})--(\ref{eq:momentump2}) is a valid approximation of the Euler equations for the prescribed asymptotic regime, it is not of much practical use due to the temporal derivatives of the wall deviations in the momentum equation (\ref{eq:momentump2}). For this reason we proceed with further simplifications using low-order approximation for the velocity $w$ and the deviation of the vessel wall.

From the equations (\ref{eq:massp2})--(\ref{eq:momentump2}) we observe that
\begin{equation}
w_t=-[\beta\eta]_x+O(\varepsilon,\delta^2), \qquad \eta_t=-r^w_x w-\frac{r^w}{2}\;w_x+O(\varepsilon,\delta^2)\ .
\end{equation}
Substituting these low-order approximations into (\ref{eq:momentump2}) we obtain the simplified momentum equation
\begin{align}\label{eq:momentump3}
[1-\delta^2\alpha {r_0}_{xx}]w_t&+[\beta\eta]_x+\delta^2\frac{(3\alpha+r_0){r_0}_x}{2}\;[\beta\eta]_{xx}+\varepsilon ww_x-\delta^2\frac{(2\alpha+r_0)r_0}{4}\;w_{xxt}+\nonumber \\
&+\varepsilon \kappa w-\delta^2\gamma [\beta( {r_0}_x w+\frac{r_0}{2}\;w_x)]_x=O(\delta^4,\varepsilon\delta^2)\ .
\end{align}

The system (\ref{eq:massp2})--(\ref{eq:momentump3}) can be the base to other more amenable Boussinesq systems along the lines of \cite{BCS2002}. In the next section we derive a simplified Boussinesq systems with favourable properties in analogy to the classical Boussinesq system for fluid flow in purely elastic vessels derived in \cite{MDL2017}.

\section{The classical Boussinesq system}\label{sec:cbsys}

In this section we proceed with some improvements on system (\ref{eq:massp2})-(\ref{eq:momentump3}) based on the evaluation of the horizontal velocity at any level of radius $r$. Specifically, from (\ref{eq:potapr}) we have that $u(x,r,t)=w(x,t)-\delta^2 \frac{r^2}{4}\;w_{xx}(x,t)+O(\delta^4)$. From this we observe that considering the fluid velocity at any radius $r=\theta r^w$ for $0\leq \theta\leq 1$ we have $u^\theta(x,t)\doteq u(x,\theta r^w,t)=w(x,t)-\delta^2 \theta^2 \frac{{r_0}^2}{4}\;w(x,t)+O(\varepsilon\delta^2,\delta^4)$ and thus
\begin{equation}\label{eq:generalu}
w(x,t)=u^\theta(x,t)+\delta^2\theta^2 \frac{{r_0}^2}{4}\; u^\theta_{xx}(x,t)+O(\varepsilon\delta^2,\delta^4)\ .
\end{equation}
Substitution of (\ref{eq:generalu}) into (\ref{eq:massp2}) and (\ref{eq:momentump3})  leads to the more general Boussinesq system:
\begin{equation}\label{eq:massp4}
\eta_t+r^w_x u^\theta+\frac{r^w}{2}\;u^\theta_x+\delta^2 \frac{(2\theta^2-1)r_0{r_0}_x}{4}\;u^\theta_{xx}+\delta^2 \frac{(2\theta^2-1)r_0^3}{16}\; u^\theta_{xxx}=O(\varepsilon\delta^2,\delta^4)\ .
\end{equation}
\begin{align}
[1-\delta^2\alpha {r_0}_{xx}]u^\theta_t&+[\beta\eta]_x+\delta^2\frac{(3\alpha+r_0){r_0}_x}{2}\;[\beta\eta]_{xx}+\varepsilon u^\theta u^\theta_x-\delta^2\frac{[2\alpha-(\theta^2-1)r_0]r_0}{4}\;u^\theta_{xxt}+\nonumber \\
&+\varepsilon \kappa u^\theta-\delta^2\gamma [\beta( {r_0}_x u^\theta+\frac{r_0}{2}\;u^\theta_x)]_x=O(\delta^4,\varepsilon\delta^2)\ .\label{eq:momentump4}
\end{align}
If we take $\gamma=0$ in the system (\ref{eq:massp4})--(\ref{eq:momentump4}) then we recover the Boussinesq system of KdV-BBM type of \cite{MDL2017}. 

The linear dispersion properties of the system (\ref{eq:massp4})--(\ref{eq:momentump4}) depend on the choice of the parameter $\theta$. Taking $\theta^2=1/2$ the resulting system has the simplest form since the third order derivative term in (\ref{eq:massp4}) is canceled, and moreover, its linear dispersion relation is very close to the linear dispersion relation of the Euler equations \cite{MDL2017}. Specifically, after taking $\theta^2=1/2$ and rearranging terms, the system  (\ref{eq:massp4})--(\ref{eq:momentump4}) is simplified to the classical Boussinesq system for viscoelastic vessels:
\begin{align}
& \eta_t+\frac{1}{2}\;(r_0+\varepsilon\eta)u_x +(r_0+\varepsilon\eta)_x u = O(\varepsilon\delta^2,\delta^4)\ , \label{eq:massp5}\\
& [1-\delta^2\alpha {r_0}_{xx}]u_t+[\beta\eta]_x+\varepsilon u u_x-\delta^2\frac{(4\alpha+r_0)r_0}{8}\;u_{xxt}+ \label{eq:momentump5} \\
&+\delta^2\frac{(3\alpha+r_0){r_0}_x}{2}\;[\beta\eta]_{xx}+\varepsilon \kappa u-\delta^2\gamma [\beta( {r_0}_x u+\frac{r_0}{2}\;u_x)]_x=O(\varepsilon\delta^2,\delta^4)\ , \nonumber
\end{align}
where we drop the $\theta$ in this notation by taking $u=u^\theta$ for $\theta=1/2$. This system is very similar to the Peregrine system of water wave theory and we usually call it the classical Boussinesq system \cite{Peregrine1967}. The classical Boussinesq system (\ref{eq:massp5})--(\ref{eq:momentump5}) after discarding the negligible terms on the right side, can be written in dimensional variables form
\begin{align}
& \eta_t+\frac{1}{2}\;(r_0+\eta)u_x +(r_0+\eta)_x u= 0\ , \label{eq:massp6}\\
& [1-\bar{\alpha} {r_0}_{xx}]u_t+[\bar{\beta}\eta]_x+ u u_x-\frac{(4\bar{\alpha}+r_0)r_0}{8}\;u_{xxt}+ \label{eq:momentump6} \\
&+\frac{(3\bar{\alpha}+r_0){r_0}_x}{2}\;[\bar{\beta}\eta]_{xx}+ \kappa u-\gamma [\bar{\beta}( {r_0}_x u+\frac{r_0}{2}\;u_x)]_x=0\ ,\nonumber
\end{align}
where 
$$
  \bar{\alpha}=\frac{\rho^w h}{\rho} \mbox{ and } \bar{\beta}(x)=\frac{E h}{\rho r_0^2(x)}\,,
$$
while all the variables $x$, $t$, $r_0(x)$, $\eta(x,t)$ and $u(x,t)$ are in dimensional form.

We underline that the new dissipative terms $\kappa u$ and $-\gamma [\bar{\beta}( {r_0}_x u+\frac{r_0}{2}\;u_x)]_x$ are totally different in nature and in mathematical properties. Their coefficients appeared in these formulas are also important for the derivation of Windkessel (0D) models \cite{NOV2011} as they are the result of asymptotic simplifications to the Euler equations and can be specified by the wall material properties.

System (\ref{eq:massp6}) -- (\ref{eq:momentump6}) extends the classical Boussinesq system derived and analysed in \cite{MDL2017} in the case of vessels with elastic walls. Using low order corrections to the dispersive terms one can extend the whole class of the Boussinesq systems derived in \cite{MDL2017}. The extended systems will differ only by the additional viscoelastic term and because of their complexity we do not proceed with their derivation here.

Further simplifications can be achieved in the system (\ref{eq:massp6})--(\ref{eq:momentump6}) by assuming that the undisturbed radius of the vessel is constant. The resulting system takes the form
\begin{align}
& \eta_t+\frac{1}{2}\;r_0\,u_x+\frac{1}{2}\;\eta u_x+ \eta_x u = 0\ , \label{eq:massp6}\\
& u_t+\bar{\beta}\eta_x+ u u_x-\frac{(4\bar{\alpha}+r_0)r_0}{8}\;u_{xxt}+  \kappa u-\gamma \bar{\beta} \frac{r_0}{2}\;u_{xx}=0\ ,\label{eq:momentump6}
\end{align}
where $\bar{\alpha}$ and $\bar{\beta}$ are constants.

The dispersion relation $\omega\ =\ \omega\,(k)$ of the derived system can be easily computed. It is given by the following quadratic algebraic equation:
\begin{equation}\label{eq:disprel}
  \Bigl(1\ +\ \frac{\bar{\alpha}}{2}\;r_{\,0}\,k^{\,2}\ +\ \frac{r_{\,0}^{\,2}}{8}\,k^{\,2}\Bigr)\,\omega^{\,2}\ +\ \ui\,\Bigl(\kappa\ +\ \frac{\gamma\,\bar{\beta}\,r_{\,0}\,k^{\,2}}{2}\Bigr)\,\omega\ -\ \frac{\bar{\beta}\,r_{\,0}\,k^{\,2}}{2}\ =\ 0\,.
\end{equation}
Here $k$ is the wavenumber and $\omega\ =\ \omega\,(k)$ is the corresponding wave frequency. The dispersion relation \eqref{eq:disprel} can be solved explicitly for $\omega$ using the standard formulas for the roots of a quadratic equation. It will contain two branches whose expressions we omit here for brevity.

\subsection{Solitary wave solutions}

\begin{figure}[ht!]
  \centering
  \bigskip
  \includegraphics[width=0.8\columnwidth]{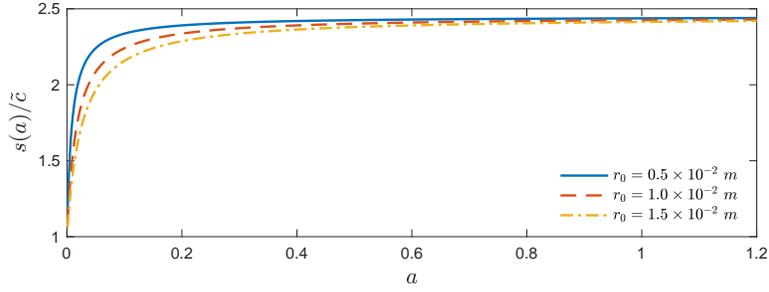}
  \caption{\small\em Speed-amplitude relation \eqref{eq:dpdampl} for solitary waves of the classical Boussinesq system for various values of $r_0$.}
  \label{fig:sa}
\end{figure}

In the case where all the dissipative terms are neglected (i.e. when $\kappa=\gamma=0$) the system possesses classical solitary wave solutions that satisfy the following speed-amplitude relation \cite{MDL2017}
\begin{equation}\label{eq:dpdampl}
s=\sqrt{6\bar{\beta}r_0\frac{ (2-\zeta)  - 2 \sqrt{ 1-\zeta}}{ \zeta^2(3-\zeta)}}\ ,
\end{equation}
where $\zeta=1-r_0^2/(a+r_0)^2$, and $s$ denotes the speed and $a$ the amplitude of the solitary wave. The graph of the speed-amplitude relationship for values of the parameters $\bar{\alpha}$ and $\bar{\beta}$ for a typical blood vessels of different radius $r_0$ is presented in Figure~\ref{fig:sa}. We observe that there is as the amplitude of the solitary waves grows their speed tend to approach an upper bound. The upper bound for the solitary wave speed can be computed by taking the limit $a\to \infty$ in (\ref{eq:dpdampl}), and which is $\sqrt{3\bar{b}r_0}$.

For the numerical generation of these solitary waves we refer to \cite{MDL2017}. Other more general systems similar to the systems derived in \cite{MDL2017} can also be derived but since they only differ on the viscoelastic term we omit their derivation here and we refer to \cite{MDL2017} for more information.

\subsection{Symmetries and conservation laws}

In the presence of any dissipative term (i.e. $\kappa^{\,2}\ +\ \gamma^{\,2}\ \neq\ 0$), the system \eqref{eq:massp6}, \eqref{eq:momentump6} possesses only two point symmetries: translations in time and in space. If both dissipative terms are neglected (i.e. $\kappa^{\,2}\ +\ \gamma^{\,2}\ =\ 0$), then we gain an additional point symmetry transformation whose infinitesimal generator is given by
\begin{equation*}
  \X\ =\ t\;\pd{}{t}\ -\ 2\,(\eta\ +\ r_{\,0})\;\pd{}{\eta}\ -\ u\;\pd{}{u}\,.
\end{equation*}
The corresponding symmetry transformation can be readily computed:
\begin{align*}
  x\ &\leftarrow\ x\,, \\
  t\ &\leftarrow\ \ue^{c}\,t\,, \\
  \eta\ &\leftarrow\ (r_{\,0}\ +\ \eta)\,\ue^{\,-\,2\,c}\ -\ r_{\,0}\,, \\
  u\ &\leftarrow\ \ue^{\,-\,c}\,u\, ,
\end{align*}
where $c$ is an arbitrary constant.

For arbitrary $\kappa$ and $\gamma$ system \eqref{eq:massp6}, \eqref{eq:momentump6} can be written in an elegant conservative form:
\begin{equation*}
  \bigl(\eta\,(\eta\ +\ 2\,r_{\,0})\bigr)_{\,t}\ +\ \bigl[\,(r_{\,0}\ +\ \eta)^{\,2}\,u\,\bigr]_{\,x}\ =\ 0\,,
\end{equation*}
\begin{multline*}
  \Bigl(\bigl(24\,u\ -\ r_{\,0}\,(r_{\,0}\ +\ 4\,\bar{\alpha})\,u_{\,x\,x}\bigr)\,\ue^{\,\kappa\,t}\Bigr)_{\,t}\ +\\ 
  \Bigl[\,\bigl(12\,u^{\,2}\ +\ 24\,\bar{\beta}\,\eta\ +\ r_{\,0}^{\,2}\,\kappa\,u_{\,x}\ +\ 4\,r_{\,0}\,\bar{\alpha}\,\kappa\,u_{\,x}\\ 
  -\ 2\,r_{\,0}^{\,2}\,u_{\,t\,x}\ -\ 8\,r_{\,0}\,\alpha\,u_{\,t\,x}\ -\ 12\,\gamma\,\bar{\beta}\,u_{\,x}\bigr)\,\ue^{\,\kappa\,t}\,\Bigr]_{\,x}\ =\ 0\,.
\end{multline*}
For instance, these conserved quantities can be used in numerical simulations to check method accuracy.

\section{Unidirectional models}\label{sec:unidir}

In this section we present unidirectional models, namely the BBM and KdV equations by considering waves that propagate mainly in one direction. In order to derive such models we consider  the following dimensionless variables:
\begin{equation}\label{eq:ndv2}
\eta^\ast=\frac{\eta}{a},~ x^{\ast}=\frac{x}{\lambda},~ r^{\ast}=\frac{r}{r_0},~  t^{\ast}=\frac{t}{T} \ ,u^\ast=\frac{u}{c_0}\ ,
\end{equation}
where here $c_0=\frac{a}{r_0}\sqrt{\frac{2Eh}{\rho r_0}}$ is a modified Moens-Korteweg characteristic speed and $T=2\frac{a\lambda}{r_0 c_0}\,$. The system (\ref{eq:massp6})--(\ref{eq:momentump6}) in dimensionless variables then is written:
\begin{align}
& \eta^\ast_{t^\ast}+u^{\ast}_{x^\ast}+\varepsilon \eta^\ast u^{\ast}_{x^{\ast}}+ 2\varepsilon \eta^\ast_{x^\ast} u^\ast = 0\ , \label{eq:massp7}\\
& u^\ast_{t^\ast}+\eta^\ast_{x^\ast}+ u^{\ast} u^{\ast}_{x^\ast}-\delta^2\frac{4\alpha^\ast+1}{8}\;u^\ast_{x^\ast x^\ast t^\ast}+ \varepsilon \kappa^\ast u^\ast-\delta^2 \gamma^\ast u_{xx}=0\ ,\label{eq:momentump7}
\end{align}
where here $\kappa^\ast=2\frac{\kappa}{\sigma}$, $\gamma^\ast=\frac{1}{2}\frac{\sigma}{\varepsilon\delta^2}\gamma$, $\sigma=\frac{c_0}{\lambda}$ and $\varepsilon=\frac{a}{r_0}$, $\delta=\frac{r_0}{\lambda}$.
Considering a flow mainly towards one direction, we can use the low-order approximation for unidirectional wave propagation \cite{Whitham} 
\begin{equation}\label{eq:unid}
u^\ast=\eta^\ast+\varepsilon F+\delta^2G+O(\varepsilon^2,\varepsilon\delta^2,\delta^4)\ ,
\end{equation}
where $F$ and $G$ are unknown functions of $x$, $t$. Substitution of (\ref{eq:unid}) into (\ref{eq:massp7})--(\ref{eq:momentump7}) gives the equations:
\begin{align}
&\eta^\ast_{t^\ast}+\eta^\ast_{x^\ast}+\varepsilon(F_{x^\ast}+3\eta^\ast\eta^\ast_{x^\ast})+\delta^2 G_x=O(\varepsilon^2,\varepsilon\delta^2,\delta^4), \label{eq:gg1}\\
&\eta^\ast_{t^\ast}+\eta^\ast_{x^\ast}+\varepsilon(F_{t^\ast}+2\eta^\ast\eta^\ast_{x^\ast}+k^\ast\eta^\ast)+\delta^2(G_{t^\ast}-\frac{4\alpha^\ast+1}{8}\;\eta^{\ast}_{x^\ast x^\ast t^\ast}-\gamma^\ast \eta^\ast_{x^\ast x^\ast})=O(\varepsilon^2,\varepsilon\delta^2,\delta^4)\ . \label{eq:gg2}
\end{align}
Choosing appropriate functions $F$ and $G$
\begin{equation}
F=-\frac{1}{4}\;{\eta^\ast}^2+\frac{\kappa^\ast}{2}\; \int\eta^\ast\ dx^\ast, \qquad G=-\frac{4\alpha^\ast+1}{16}\;\eta^\ast_{x^\ast t^\ast}-\frac{\gamma^\ast}{2}\;\eta^\ast_{x^\ast}~ ,
\end{equation} equations (\ref{eq:gg1}) and (\ref{eq:gg2}) coincide up to the order $O(\varepsilon,\delta^2)$ to a single equation for $\eta^{\ast}$, namely,
the dimensionless BBM equation:
\begin{equation}\label{eq:bbm}
\eta^\ast_{t^\ast}+\eta^\ast_{x^\ast}+\varepsilon \frac{5}{2}\;\eta^\ast\eta^\ast_{x^\ast}-\delta^2\frac{4\alpha^\ast+1}{16}\;\eta^\ast_{x^\ast x^\ast t^\ast} +\varepsilon\frac{\kappa^\ast}{2}\;\eta-\delta^2\frac{\gamma^\ast}{2}\;\eta^\ast_{x^\ast x^\ast}=0\ .
\end{equation}
In dimensional variables (\ref{eq:bbm}) takes the form
\begin{equation}\label{eq:bbmd}
\eta_t+\tilde{c}\eta_x+ \frac{5}{2}\;\frac{1}{r_0}\;\tilde{c}\eta\eta_x-\frac{(4\bar{\alpha}+r_0)r_0}{16}\;\eta_{xxt} +\frac{\kappa}{2}\;\eta-\frac{\tilde{c}}{2}\;\gamma\eta_{xx}=0\ ,
\end{equation}
where here $\tilde{c}=\sqrt{\frac{Eh}{2\rho r_0}}$ is the standard Moens-Korteweg characteristic speed. The dispersion relation $\omega\ =\ \omega\,(k)$ can be easily computed:
\begin{equation}\label{eq:disprelbbm}
  \omega\ =\ \frac{\tilde{c}\,k}{1\ +\ \dfrac{r_{\,0}\,(4\,\bar{\alpha}\ +\ r_{\,0})}{16}\,k^{\,2}}\ -\ \frac{\ui}{2}\;\frac{\kappa\ +\ \gamma\,\tilde{c}\,k^{\,2}}{1\ +\ \dfrac{r_{\,0}\,(4\,\bar{\alpha}\ +\ r_{\,0})}{16}\,k^{\,2}}\,.
\end{equation}

In the absence of any form of dissipation, it is known that the BBM equation possesses classical solitary waves propagating with speed $c_s$ given by the formula, \cite{MDL2017},
\begin{equation}\label{eq.sw}
\eta(x,t) = 3\frac{c_s-a}{b}\;{\rm sech}^2\left(\sqrt{\frac{c_s-a}{4c_s c}}\;(x-c_s t)\right)\ ,
\end{equation}
with $a=\tilde{c}$, $b=5\tilde{c}/2r_0$, and $c=\tilde{c}(4\bar{\alpha}+r_0)r_0/16$.

Observing that $\eta^\ast_{x^\ast}=-\eta^\ast_{t^\ast}+O(\varepsilon,\delta^2)$ from (\ref{eq:bbm}) and modifying accordingly the dispersive term of the BBM equation we obtain the analogous KdV equation:
\begin{equation}\label{eq:KdV}
\eta^\ast_{t^\ast}+\eta^\ast_{x^\ast}+\varepsilon \frac{5}{2}\;\eta^\ast\eta^\ast_{x^\ast}+\delta^2\frac{4\alpha^\ast+1}{16}\;\eta^\ast_{x^\ast x^\ast x^\ast} +\varepsilon\frac{\kappa^\ast}{2}\;\eta-\delta^2\frac{\gamma^\ast}{2}\;\eta^\ast_{x^\ast x^\ast}=0\ ,
\end{equation}
which in dimensional form becomes
\begin{equation}\label{eq:KdV2}
  \eta_t+\tilde{c}\eta_x+ \frac{5}{2}\;\frac{1}{r_0}\;\tilde{c}\eta\eta_x+\tilde{c}\frac{(4\bar{\alpha}+r_0)r_0}{16}\;\eta_{xxx} +\frac{\kappa}{2}\;\eta-\frac{\tilde{c}}{2}\;\gamma\eta_{xx}=0\ .
\end{equation}
The dispersion relation $\omega\ =\ \omega\,(k)$ of the derived KdV equation \eqref{eq:KdV2} can be easily computed:
\begin{equation}\label{eq:disprelkdv}
  \omega\ =\ \tilde{c}\,k\ -\ \frac{r_{\,0}\,\tilde{c}\,(4\,\bar{\alpha}\ +\ r_{\,0})}{16}\;k^{\,3}\ -\ \frac{\ui}{2}\;(\kappa\ +\ \gamma\,\tilde{c}\,k^{\,2})\,.
\end{equation}
The imaginary part $\Im\omega\,(k)$ comes with the negative sign, which indicates that we have effectively introduced dissipation into the model.

\subsection{Symmetries and conservation laws}

Similar to the Boussinesq-type model \eqref{eq:massp6}, \eqref{eq:momentump6}, the BBM equation \eqref{eq:bbmd} possesses only space and time translations symmetries when $\kappa^{\,2}\ +\ \gamma^{\,2}\ \neq\ 0\,$. However, when we neglect completely the dissipation (i.e. $\kappa^{\,2}\ +\ \gamma^{\,2}\ =\ 0$), another symmetry appears with the following infinitesimal generator:
\begin{equation*}
  \X\ =\ t\;\pd{}{t}\ -\ \Bigl(\frac{2}{5}\;r_{\,0}\ +\ \eta\Bigr)\;\pd{}{\eta}\,.
\end{equation*}
The corresponding symmetry transformation can be readily computed:
\begin{align*}
  x\ &\leftarrow\ x\,, \\
  t\ &\leftarrow\ \ue^{c}\,t\,, \\
  \eta\ &\leftarrow\ \Bigl(\,\frac{2}{5}\;r_{\,0}\ +\ \eta\,\Bigr)\,\ue^{\,-\,c}\ -\ \frac{2}{5}\;r_{\,0}\, ,
\end{align*}
where $c$ is again an arbitrary constant. The BBM equation \eqref{eq:bbmd} admits also the following conservative form:
\begin{multline*}
  \Bigl(\bigl(\,\eta\ -\ \frac{r_{\,0}\,(r_{\,0}\ +\ 4\,\bar{\alpha})}{48}\;\eta_{\,x\,x}\,\bigr)\,\ue^{\frac{\kappa}{2}\;t}\Bigr)_{\,t}\ +\\ 
  \Bigl[\,\frac{1}{2\,r_{\,0}}\;\Bigl(\,\frac{5}{2}\;\tilde{c}\,\eta^{\,2}\ +\ 2\,\tilde{c}\,r_{\,0}\,\eta\ +\ \frac{r_{\,0}}{48}\;\bigl(r_{\,0}^{\,2}\,\kappa\ +\ 4\,\kappa\,\bar{\alpha}\,r_{\,0}\ -\ 48\,\tilde{c}\,\gamma\bigr)\,\eta_{\,x}\\
   -\ \frac{r_{\,0}^{\,2}\,(r_{\,0}\ +\ 4\,\bar{\alpha})}{12}\;\eta_{\,x\,t}\,\Bigr)\,\ue^{\frac{\kappa}{2}\;t}\,\Bigr]_{\,x}\ =\ 0\,.
\end{multline*}

The point symmetry group of the KdV equation \eqref{eq:KdV2} has one extra symmetry transformation even in the dissipative case (i.e. $\kappa^{\,2}\ +\ \gamma^{\,2}\ \neq\ 0$). It is given by the following infinitesimal generator:
\begin{equation*}
  \X\ =\ \ue^{-\,\frac{\kappa}{2}\;t}\;\pd{}{x}\ -\ \frac{\kappa\,r_{\,0}}{5\,\tilde{c}}\,\ue^{-\,\frac{\kappa}{2}\;t}\;\pd{}{\eta}\,.
\end{equation*}
The corresponding symmetry transformation can be readily obtained:
\begin{align*}
  x\ &\leftarrow\ x\ +\ c\,\ue^{-\,\frac{\kappa}{2}\;t}\,\,, \\
  t\ &\leftarrow\ t\,, \\
  \eta\ &\leftarrow\ \eta\ -\ \frac{r_{\,0}\,\kappa}{5\,\tilde{c}}\;c\,\ue^{-\,\frac{\kappa}{2}\;t}\,,
\end{align*}
with $c$ an arbitrary constant. 
If we neglect the dissipative effects in KdV equation \eqref{eq:KdV2} (i.e. $\kappa^{\,2}\ +\ \gamma^{\,2}\ =\ 0$), there are two additional point symmetry transformations (we always keep time and space translations) given by infinitesimal generators:
\begin{align*}
  \X_{\,3}\ &=\ t\;\pd{}{x}\ +\ \frac{2\,r_{\,0}}{5\,\tilde{c}}\;\pd{}{\eta}\,, \\
  \X_{\,4}\ &=\ \frac{x}{3}\;\pd{}{x}\ +\ t\;\pd{}{t}\ -\ \Bigl(\frac{2}{3}\;\eta\ +\ \frac{4\,r_{\,0}}{15}\Bigr)\;\pd{}{\eta}\, .
\end{align*}
The corresponding transformations can be readily computed:
\begin{align*}
  x\ &\leftarrow\ x\,, &
  x\ &\leftarrow\ x\,\ue^{\,\frac{c}{3}}\,, \\
  t\ &\leftarrow\ t\,\ue^{\,c}\,, & 
  t\ &\leftarrow\ t\,\ue^{\,c}\,, \\
  \eta\ &\leftarrow\ \Bigl(\,\eta\ +\ \frac{2\,r_{\,0}}{5}\,\Bigr)\,\ue^{\,-\,c}\ -\ \frac{2\,r_{\,0}}{5}\,, &
  \eta\ &\leftarrow\ \Bigl(\,\eta\ +\ \frac{2\,r_{\,0}}{5}\,\Bigr)\,\ue^{\,-\,\frac{2}{3}\;c}\ -\ \frac{2\,r_{\,0}}{5}\, ,
\end{align*}
where $c$ is an arbitrary constant. 

The KdV equation \eqref{eq:KdV2} (with dissipative terms) admits the following conservative form:
\begin{multline*}
  \Bigl(\eta\,\ue^{\frac{\kappa}{2}\;t}\Bigr)_{\,t}\ +\ \Bigl[\,\tilde{c}\Bigl(\eta\ +\ \frac{5}{4\,r_{\,0}}\,\eta^{\,2}\ -\ \frac{\gamma}{2}\;\eta_{\,x}\ +\ \frac{r_{\,0}\,(4\,\bar{\alpha}\ +\ r_{\,0})}{16}\;\eta_{\,x\,x}\Bigr)\,\ue^{\frac{\kappa}{2}\;t}\,\Bigr]_{\,x}\ =\ 0\,.
\end{multline*}

In the next Section we compare the BBM equation with the Boussinesq system and we study the dissipation effects due to the viscosity of the fluid and the viscoelastic walls.
 
\section{The effect of viscoelasticity}

In this section we study the effects of the viscoelasticity and the dissipation due to the viscous nature of a fluid in the propagation of solitary and periodic waves and compare the BBM equation with the classical Boussinesq system. 

\subsection{Dissipative effects on solitary waves}

We consider the system (\ref{eq:massp6})--(\ref{eq:momentump6}) for a vessel with undisturbed radius $r=0.01~m$, wall thickness $h=0.0003~m$, Young modulus $E=4.1\times 10^5~kg/m\cdot s^2$, wall density $\rho^w=1000~kg/m^3$, fluid density $\rho=1060~kg/m^3$ and length $0.4~m$ for the propagation of a solitary wave with amplitude $a= 0.0035~m$ obtained using the Petviashvili method for the classical Boussinesq system described in \cite{MDL2017}. For the solitary waves of the BBM equation we used the exact formula (\ref{eq.sw}). We discretised both models using the standard pseudo-spectral method in space and the fourth-order, four-stage, classical Runge-Kutta scheme for the integration in time. Although the particular case is far from being realistic, the parameters are chosen to resemble a large blood vessel and it serves only the purposes of the study of the dissipative effects of the new equations. 

\begin{figure}[ht!]
  \centering
  \includegraphics[width=0.8\columnwidth]{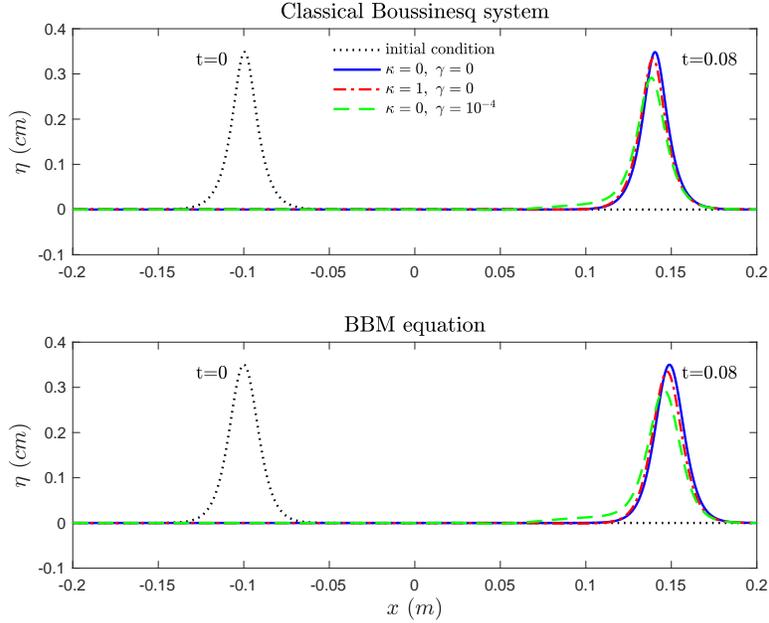}
  \caption{\small\em The effects of viscoelastic walls and viscosity of the fluid in the propagation of a solitary wave. ($r=0.01~m$, $h=0.0003~m$, $E=4.1\times 10^5~kg/m\cdot s^2$, $\rho^w=1000~kg/m^3$, $\rho=1060~kg/m^3$).}
  \label{fig1}
\end{figure}

In order to study the effects of the dissipation and viscoelasticity of the vessel wall on the propagation of the solitary wave we consider three cases: (i) a vessel with elastic walls but with no viscosity and an inviscid fluid ($\kappa=0,~\gamma=0$); (ii) a vessel with elastic walls but with no viscosity ($\gamma=0$) and a viscous fluid ($\kappa=1~s^{-1}$); and (iii) a vessel with viscoelastic walls ($\gamma=10^{-4}~s$) and an inviscid fluid ($\kappa=0$). We integrate the system numerically until the maximum time $t=0.08~s$. The results obtained at time $t=0.08~s$ are presented in Figure \ref{fig1}.

The amplitude of the solitary wave was reduced approximately by $6\%$ when $\kappa=1$ and $\gamma=0$ and by $17\%$ when $\kappa=0$ and $\gamma=10^{-4}$ for both models. We also observe that although the initial solitary waves have slightly different shape and speed, the shape of the solutions at $t=0.08~s$ is very similar in both cases. 

In order to complete the study of the importance of the new dissipative terms we also performed an experiment combining the two dissipative terms with $\kappa=1$ and $\gamma=10^{-4}$. The amplitude of the solitary wave in this case was reduced by $20\%$. It seems that the effects from the viscoelastic walls are very important and should be included in future studies. Moreover, we observe that small amplitude waves propagate in the opposite direction with respect to the direction of the propagation of the solitary wave. For the accurate description of waves propagating in different directions it is required the model to be able to describe two-way propagation of the waves, and the new models are capable of doing that. On the contrary unidirectional models such as the KdV or BBM equations although they require the knowledge of only one quantity (namely the initial condition for the initial disturbance of the wall or the pressure), they have certain disadvantages, especially if it is required to consider reflections due to branching or due to other forms of obstacles.

Although the propagation of a solitary wave in a dissipative environment is interesting for the investigation of the effects of the viscosity and viscoelasticity, it is even interesting to examine the interaction of two solitary waves in the same environment. Even if the interaction of two solitary waves is governed by the nonlinear properties of the model, the linear dissipative terms can affect the interaction quite dramatically. 

\begin{figure}[ht!]
  \centering
  \includegraphics[width=\columnwidth]{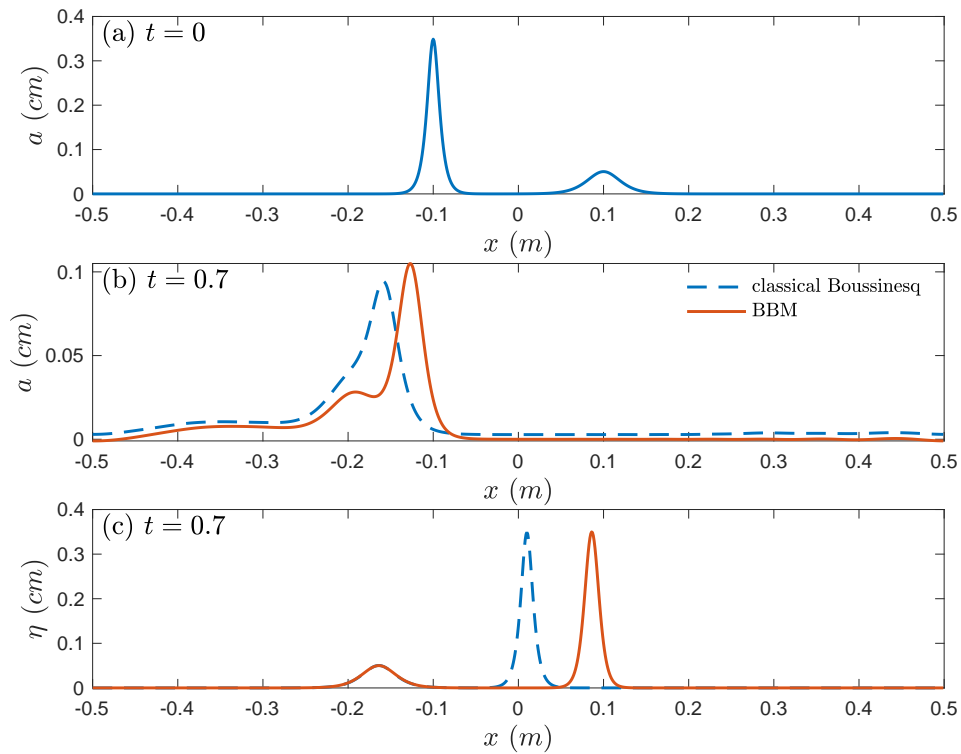}
  \caption{\small\em The effects of viscoelastic walls and viscosity of the fluid in the interaction of two solitary waves at $t=0.7~sec$. ($r=0.01~m$, $h=0.0003~m$, $E=4.1\times 10^5~kg/m\cdot s^2$, $\rho^w=1000~kg/m^3$, $\rho=1060~kg/m^3$).}
  \label{fig3}
\end{figure}
\begin{figure}[ht!]
  \centering
  \includegraphics[width=\columnwidth]{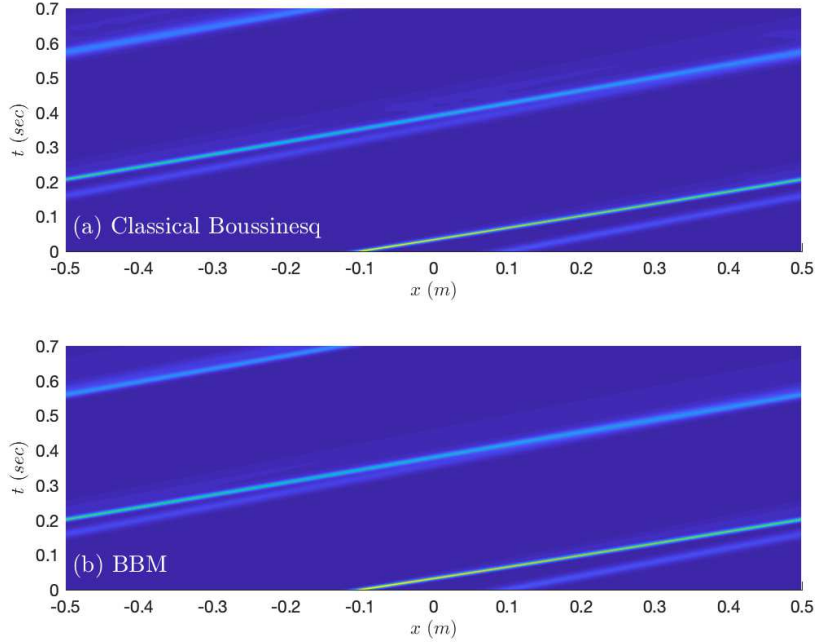}
  \caption{\small\em The effects of viscoelastic walls and viscosity of the fluid in the interaction of two solitary waves at $t=0.7~sec$. ($r=0.01~m$, $h=0.0003~m$, $E=4.1\times 10^5~kg/m\cdot s^2$, $\rho^w=1000~kg/m^3$, $\rho=1060~kg/m^3$).}
  \label{fig4}
\end{figure}

In the next experiment we consider the interaction of two solitary waves with speeds amplitudes $0.0035~m$ and $0.0005~m$. All the parameters of the equations are the same as before whereas $\kappa=1$ and $\gamma=10^{-4}$. For this we take the spatial interval $[-0.5,0.5]$ where the solitary waves are located at $x=-0.1$ and $x=0.1$ respectively. Although the solitary waves of the BBM equation and the classical Boussinesq system have similar shapes they differ slightly in their speeds. For this reason certain discrepancies related to the location of the solitary waves after the interaction can be observed. Figure \ref{fig3} (a) depicts the initial location of the two solitary waves. In Figure \ref{fig3} (b) the solution at $t=0.7~sec$ is presented in the presence of dissipation, and Figure \ref{fig3} (c) the evolution of the same initial conditions without the presence of any dissipation is presented at the same time $t$. The dissipation not only affect the amplitudes of the solitary waves but also the length of the interaction. Because the solitary waves of the BBM equation propagate faster than the corresponding solitary waves of the classical Boussinesq system we observe that at the same time $t=0.7~sec$ the interaction of the two solitary waves of the BBM equation is a  more advanced stage of the separation phase compared to the analogous results of the classical Boussinesq system. Figure \ref{fig4} shows a phase diagram of the two solitary waves during the interaction for both systems.

\subsection{Dissipative effects on periodic waves}

In order to further assess the effects of the dissipation due to the viscoelastic walls one can examine the linear dispersion relations of the equations at hand. All the dispersion relations (\ref{eq:disprel}), (\ref{eq:disprelbbm}) and (\ref{eq:disprelkdv}) have a nonzero imaginary part due to the dissipative terms. Thus, plane waves of the form $\eta(x,t)=A_0e^{i(kx-\omega(k) t)}$ will have a decaying amplitude of the form $A_0e^{\Im\omega t}$. Of course the new dissipative equations are all non-linear. In fact in this section we show that the dissipation rate estimate we obtain from the linear dispersion relation of the models predict very accurately the actual dissipation for the nonlinear equations. For the purposes of this experiment we consider $\kappa=1$ and $\gamma=10^{-4}$.  In order to study the combined effects of the nonlinearities with the dissipation we consider a simple periodic wave $\eta_0(x)=A_0\sin(kx)$ with $A_0=10^{-3}~m$ and $k=2$ for the same mathematical models and parameters as before.
\begin{figure}[ht!]
  \centering
  \includegraphics[width=0.8\columnwidth]{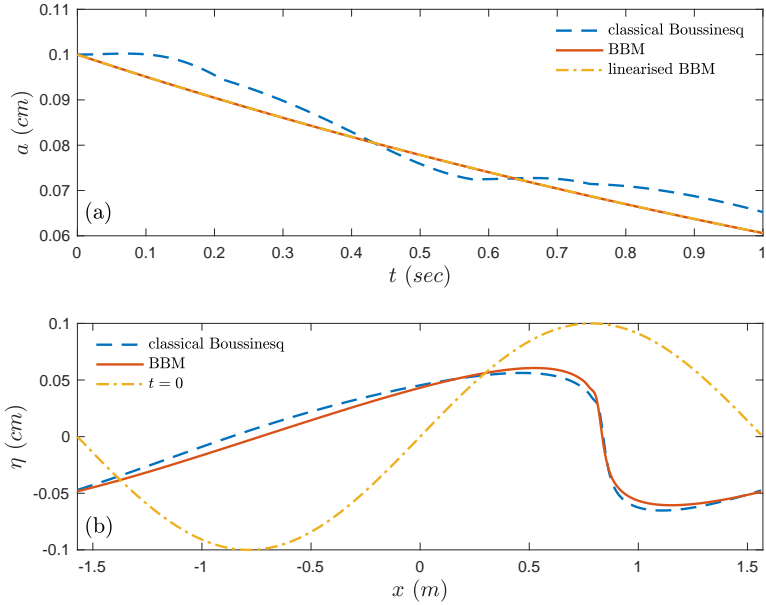}
  \caption{\small\em The effects of viscoelastic walls and viscosity of the fluid in the propagation of a periodic wave at $t=1~sec$. ($r=0.01~m$, $h=0.0003~m$, $E=4.1\times 10^5~kg/m\cdot s^2$, $\rho^w=1000~kg/m^3$, $\rho=1060~kg/m^3$).}
  \label{fig2}
\end{figure}
Figure \ref{fig2} (a) shows the amplitude of the periodic wave as a function of time for both cases, namely, the classical Boussinesq and the BBM equations. For the classical Boussinesq system an approximate velocity profile was chosen to simulate one-way propagation of the periodic wave. Because this formula is not exact we observe discrepancies between the computed and the theoretical amplitude of the classical Boussinesq system. The predicted amplitude from the linear theory apparently is almost identical to the numerical amplitude of the nonlinear BBM equation. This fact shows that dissipative effects are linear processes and thus can be included in other derived 0D linearized models. For the damping rate of dissipative KdV equations we refer to \cite{CS2013}. Figure \ref{fig2} (b) presents the profile of the $\eta$-solution at $t=1~sec$. We observe that although the initial condition for the $u$-component of the classical Boussinesq solution is not exact, the profiles are very similar while the effect of dissipation is very strong. It is noted that the pulses have complete one period up to $t=1~sec$. It is worth to mention that the wave profile will break in later time generating dispersive shock waves as it is well-known \cite{Whitham}.

\section{Conclusions} 

In this paper we derived new weakly nonlinear and weakly dispersive asymptotic equations that describe the irrotational and dissipative flow of a fluid in pipes with viscoelastic walls. We also derived unidirectional equations of BBM and KdV type when the undisturbed radius is constant along the pipe. In order to study the dissipative effects due to fluid viscosity and the viscoelastic walls, we considered solitary and periodic waves propagating in a vessel of constant undisturbed radius and with parameters that resemble a large blood vessel. We observed that the dissipation caused by the viscoelastic wall is equally important compared to the dissipation caused by the viscosity of the fluid or more important, and therefore should not be neglected. It is also observed that the dissipative effects can be described very accurately by linear approximations. The new asymptotic models have the potential to contribute in the derivation of new lumped parameter models that can be used in operational situations where measurements of the pressure and flow of the fluid are required.

\section*{Acknowledgments}
The work of D. Mitsotakis was supported by the Marsden Fund administered by the Royal Society of New Zealand. The authors would like to thank sincerely the anonymous referees who helped us to improve this paper with their comments and suggestions.

\bibliographystyle{plain}
\bibliography{biblio}
\bigskip

\end{document}